\documentclass{sig-alternate}
\usepackage[ruled,vlined,noresetcount]{algorithm2e}
\usepackage{graphicx}
\usepackage[justification=centering]{caption}
\usepackage[noadjust]{cite}
\usepackage[utf8]{inputenc}
\usepackage[T1]{fontenc}

\begin{document}
%
\conferenceinfo{}{Lyon, France}

\title{EvoCut : A new Generalization of Albert-Barab\'asi Model for Evolution of Complex Networks}
\subtitle{}
%
%
%
%
%

\numberofauthors{6} 
%
\author{
%
%
\alignauthor
Shailesh Kumar Jaiswal\\
       \affaddr{National Institute of Technology }\\
       \affaddr{Meghalaya. 793003}\\
       \email{shaileshsr141@gmail.com}
\alignauthor
Nabajyoti Medhi\\\
       \affaddr{Tezpur} \\ 
       \affaddr{University }\\
       \affaddr{Assam - 784028}\\
       \email{nmedhi@tezu.ernet.in}
\alignauthor Manjish Pal\\
       \affaddr{National Institute of Technology }\\
       \affaddr{Meghalaya. 793003}\\
       \email{manjishster@nitm.ac.in}
\and  
\alignauthor Mridul Sahu\\
       \affaddr{National Institute of Technology }\\
       \affaddr{Meghalaya. 793003}\\
       \email{mridulsahu01@gmail.com}
\alignauthor Prashant Sahu\\
       \affaddr{National Institute of Technology }\\
       \affaddr{Meghalaya. 793003}\\
       \email{pras13hant@gmail.com}
\alignauthor Amal Dev Sarma\\
       \affaddr{National Institute of Technology }\\
       \affaddr{Meghalaya. 793003}\\
       \email{amal.sarma@nitm.ac.in}
}

\date{24 April 2018}

\maketitle
\begin{abstract}
With the evolution of social networks, the network structure shows dynamic nature in which nodes and edges appear as well as disappear for various reasons. The role of a node in the network is presented as the number of interactions it has with the other nodes. For this purpose a network is modeled as a graph where nodes represent network members and edges represent a relationship among them. Several models for evolution of social networks has been proposed till date, most widely accepted being the Barab\'asi-Albert \cite{Network science} model that is based on \emph{preferential attachment} of nodes according to the degree distribution. This model leads to generation of graphs that are called \emph{Scale Free} and the degree distribution of such graphs follow the \emph{power law}. Several generalizations of this model has also been proposed. In this paper we present a new generalization of the model and attempt to bring out its implications in real life.
\end{abstract}

\terms{WWW, Social media networks, Online social networks.}


\section{Introduction}
Our lives are surrounded with several complex phenomena which are at times very difficult to explain or understand. These phenomena usually encompass a lot of spheres of our lives. One such phenomena is how certain biological, geological, physical, astronomical, financial and social systems show a very peculiar similarity that is they all exhibit a so called \emph{Scale Free Property} \cite{Scale-free networks}. This property says that certain features of these systems follow a common pattern. For example in case of geology, most of the earthquakes that occur on this planet are nominal whereas few are gigantic, in case of financial systems most people on earth have small incomes and a few have a monumental incomes, in case of social systems, there are few celebrities who are extremely popular whereas most of the other are not and this list of examples continues on and on. As it turns out this kind of a Scale free pattern has been known to scientists and researchers for long. Herbert Simon \cite{behavioral model} showed the existence of such properties way back in the 1950s. Although this property is very common, present literature exhibits that there is not much understanding of as to why and how it occurs. In recent years with the advent of new technologies like the internet, WWW, social media etc., a new stream of research has come up which seeks to explain this phenomenon in several complex systems by employing mathematical models. Here the key is to find the existence of a graph theoretic structure that is present in these systems and show that the Scale Free Property is present in that graph. Researchers have shown that the presence of this property means that the corresponding graph follows the \emph{power law} degree distribution. The most well known among such models is the model of Barab$\acute{a}$si-Albert that describes a model based on \emph{preferential attachment}. This model ensures that as this graph evolves by the addition of new nodes, each new node gets attached to a set of $m$ existing nodes where the attaching probability of the new node is proportional to the degree of that node already present in the graph. This model although considered to be fairly satisfactory, is not very realistic at times and has several drawbacks. Due to this several generalizations of this model has been proposed that try to improvise on the properties of the graph generated. In this paper we propose a new model that is a generalization of the AB model and is based on the \emph{cuts} in the graph. Our model is very novel and has no apparent links with the already existing generalization of the Barab\'asi-Albert (BA) model. In this paper, we introduce our model, which we have named as the \emph{EvoCut} model, and describe its properties. We further bring out how this model is more realistic than the already existing models.                    

\subsection{The Scale Free Property}
As mentioned in the previous section, the scale free property \cite{Scale-free networks} basically means that the occurrence of very high is small and there is an abundance of small. In the context of networks and graphs this means that there are very few nodes with high degree and there are a lot of nodes with small degree. A more technical way of putting it is to say that the degree distribution follows the \emph{power law} i.e. $p(k) \propto k^{-\gamma}$ where $\gamma \in (2,3]$; $p(k)$ is the number of nodes with degree $k$ divided by $2m$ (which is the total degree of nodes). This alternatively means that on log scale the plot of $p(k)$ vs. $k$ is linear. 

\includegraphics[scale = 0.35]{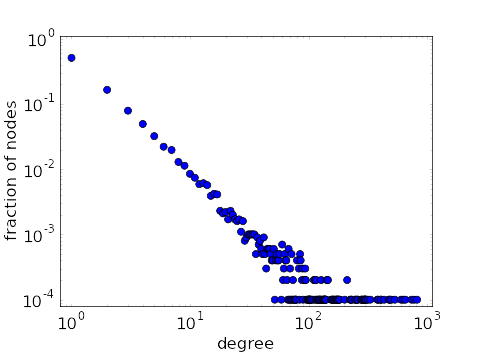}

\section{Prior Work on Generalization of Albert-Barab\'asi model}
In this section we present some prior models that are based on the generalization of the Barab$\acute{a}$si-Albert model.

\subsection{Dorogovtsev-Mendes-Samukhin (DMS) \\Model}
Dorogovtsev, Mendes and Samukhin present a model \cite{Structure}
that generalizes the BA model in the following sense, they incorporate
the initial state of the nodes in the network and the degree ($k$) is dependent on the initial state ($s$) and time ($t$). They come up with a dynamics that achieves power law when the degree is very large
i.e. $p(k) \propto k^{-\gamma}$ when $k \rightarrow \infty$ and
$\gamma \in (2,\infty)$. The average connectivity $k(s,t) \propto (s/t)^{-\beta}$ as $(s/t) \rightarrow 0$. This generalization makes the model very unrealistic because $\gamma \in (2,\infty)$ whereas in reality $\gamma \in (2,3]$, as supported by AB model.

\subsection{Antal-Krapivsky-Redner (AKR) Model}
The AKR model \cite{Dynamics} generalizes the BA model by considering a graph in which the links are associated with a notion of friendship (+1) or enmity (-1) and the notion of a node is replaces by a collection of three nodes linked with either (+1) or (-1) links, which is called a triad. The model defines  a notion of a balance which is the product of all the values of the links in a particular triad. If this product is equal to 1 then it is called \emph{balanced} otherwise it is called \emph{imbalanced}. In this model, the authors define a dynamics that tries to maintain the balance of the triads and describes how the degree of a link (which is the number of triads in which it is involved) changes with time as the network is allowed to evolve according to the dynamics. They derive a set of differential equations that governs this rate and solve them to get a distribution same as the power law.

\subsection{Sole-Pastor-Satorras-Smith-Kepler (SPSK) Model}

The Sol$\acute{e}$, Pastor-Satorras, Smith, Kepler (SPSK) model \cite{proteome evolution} uses 3 mechanisms \emph{duplicate}, \emph{divergence} and \emph{mutate} for the process of evolution. Using the operation \emph{Duplicate} one can copy a randomly selected node along with its connections, using \emph{Divergence} operation one can delete some connections made after the duplicate operation and the \emph{Mutate} operation allows us to add connections once the duplicate operation is applied. It has been shown by them that this generalization produces the \emph{power law} but leads to unrealistic assortativity and clustering. This is a generalization of the BA model because it allows more possible ways of connection of a new node with the existing graph.

\includegraphics[scale = 0.65]{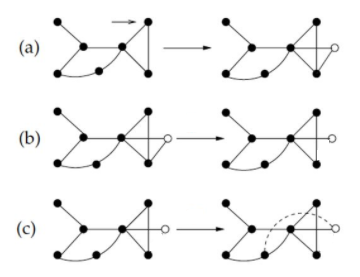}

It is customary to note that there have been some generalizations of preferential attachment model that are based on the addition and deletion of nodes \cite{Topological,Exact solutions}. 

\section{Our Contribution}
In this paper we present a new model that generalizes the BA model.
As compared to the previous models our model is a natural generalization of the BA model and is more intuitive than the previous generalizations. It uses the combinatorial properties of the given graph and does not create unrealistic links as in the case of SPSK model and DMS model. Our model has two variants, one of which generates scale free networks (as in BA model) and the other one gives rise to a family of graphs in which the degree distribution follows the stretched exponential distribution rather than the power law, which is similar to the SPSK model. In the following sections we describe in detail our proposed model and also discuss the properties of the two variants of the model.

\section{Preliminaries}
We model the network as an undirected graph in which nodes represent the network members
and an undirected edge represents a relationship between them. Initially the graph is considered to have $n_0$ number of nodes with a few links $m_0$. Our evolution process crucially uses the notion of cuts which is an important combinatorial property of a graph. Given a graph $G = (V,E)$ a subset of vertices $S$ and $\bar{S}= V\backslash S$, a \emph{Cut} is defined as the set $E(S,\bar{S})$ $= \{(a,b) | a \in S \mbox{ and } b\in \bar{S}\}$. Our evolution process uses the size of the cut crucially making it
substantially different from all the prior models that do not use the notion of cut in 
any way. We also define a $k$-neighborhood of a particular vertex $v$ as the number of nodes of $G$
which are with in a distance $k$ from the node $v$. Thus 
$B(v,k) = \{u | d(v,u) \leq k \}$. We also define the set $B'(v,k)$ as the set \\

$\{u\in B(v,k) | (u,v) \in E(B(v,k), \overline{B(v,k)}); v \in \overline{B(v,k)} \}$

\section{The EvoCut Model}
In this section we introduce our model and
describe in detail its two variants. In the first variant a new
node is attached to a node which has the maximum pulling
power which is based on the size of the $k$-neighborhood of 
that node whereas in the second variant the node is attached to
a randomly chosen node which is on the boundary of the $k$-neighborhood
set of the node with maximum pulling power.

\begin{algorithm}
\SetAlgoLined
 $G_0 = (V_0,E_0)$ ;  $t = 0$; $|V_0| = n_0$ and $E_0 = m_0$, $k = k_o$, $Y = 0$, $m = 0$\;
 \While{($n_t \leq N$)}{
  Let $v_t$ be the new node at time $t$\;
  \For{$v \in V_t$}{
  Compute $x = |E(B(v,k), \overline{B(v,k)}|$\;
  $Y = Y + x$\;
  }

  Compute $x_v = \mbox{max}_v\{\frac{|E(B(v,k), \overline{B(v,k)}|}{Y}$\}\;
  Compute $v'= \mbox{argmax}_v\{x_v\}$\;
  $E_t = E_t \bigcup (v_t,v'); V_t = V_t \bigcup v_t$\;
  $t = t+1$\; 
  $Y = 0$\;
  }
 \caption{Model A}
\end{algorithm}

\begin{algorithm}
\SetAlgoLined
 $G_0 = (V_0,E_0)$ ;  $t = 0$; $|V_0| = n_0$ and $E_0 = m_0$, $k = k_o$, $Y = 0$, $m = 0$\;
 \While{($n_t \leq N$)}{
  Let $v_t$ be the new node at time $t$\;
  \For{$v \in V_t$}{
  Compute $x = |E(B(v,k), \overline{B(v,k)}|$\;
  $Y = Y + x$\;
  }

  Compute $x_v = \mbox{max}_v\{\frac{|E(B(v,k), \overline{B(v,k)}|}{Y}$\}\;
  Select a node $v'$ randomly uniformly from the set $B'(v,k)$\;
  $E_t = E_t \bigcup (v_t,v'); V_t = V_t \bigcup v_t$\;
  $t = t+1$\;
  $Y = 0$\;
 }
 \caption{Model B}
\end{algorithm}

\section{Properties of the EvoCut Model}
In this section we describe in detail the 
properties and the reasoning behind coming up
with these models. In this model we allow the nodes to 
be attached one by one and the pulling power of 
a particular node is defined by the size of the
cut $E(B(v,k), \overline{B(v,k)}$ which the
number of edges in the $k$-neighborhood of the 
node $v$ where $k$ is a parameter between $[0-(n-1)]$.
The pulling power is same for both the models $A$ and $B$.
The other features of the models is being mentioned in
the following subsections.

\subsection{Model A - Deterministic Case} 
In this model a particular node is attached to that
node which has the maximum pulling power based on 
a given value of $k$. This model is thus fully \emph{deterministic}
i.e. doesn't use any source of randomness. The process starts with
an initial graph $G_0 = (V_0,E_0)$ and goes until the number of nodes
in the graph is less than $N$. Every time when a new node arrives,
the \textbf{for} loop computes the normalizing factor 
\begin{eqnarray}
\sum_{v}|E(B(v,k), \overline{B(v,k)}|
\end{eqnarray}
and the later part of the model computes the node $v$ that maximizes 
the ratio 
\begin{eqnarray}
\frac{B(v,k)}{\sum_{u}|E(B(u,k), \overline{B(u,k)}|}
\end{eqnarray}
and attaches the new node with this node.

\subsection{Model B - Randomized Case}
In this model a particular node is attached to a randomly chosen
node on the boundary of the set $B(v,k)$. The boundary is defined as the
nodes in $B(v,k)$ which are incident to the edges in the set 
$E(B(v,k), \overline{B(v,k)}$. This model thus uses randomness in 
choosing the node to which the incoming node should be added. As
in the previous case the model first computes the node $v$ that
maximizes the factor 
\begin{eqnarray}
\frac{B(v,k)}{\sum_{u}|E(B(u,k), \overline{B(u,k)}|}
\end{eqnarray}
and then randomly chooses a node from the set $B'(v,k)$ and
attach the new node with that node.

\subsection{As a Generalization of BA Model}
One can observe that in both the models mentioned above if we fix the parameter $k$ as 0, then we get the (BA) model. This
is simply because when $k=0$, $B(v,k) = v$ and
the set $E(B(v,k), \overline{B(v,k)}$ is equal to the degree 
of $v$ and in that case the attaching probability distribution is exactly equal to the degree distribution of the graph at time $t$ which is same as the BA model. Thus, in our model, only the special case of $k=0$ results in the BA model.

\section{Experimental Results and Analysis}
In this section we present the degree distributions that
we obtain as we let our model to evolve on a set of nodes with
certain initial condition.   

\subsection{Analysis of Model A}	 
The following are the plots that we 
obtain once the degree distributions are generated for
\emph{Model A}.

\includegraphics[scale = 0.29]{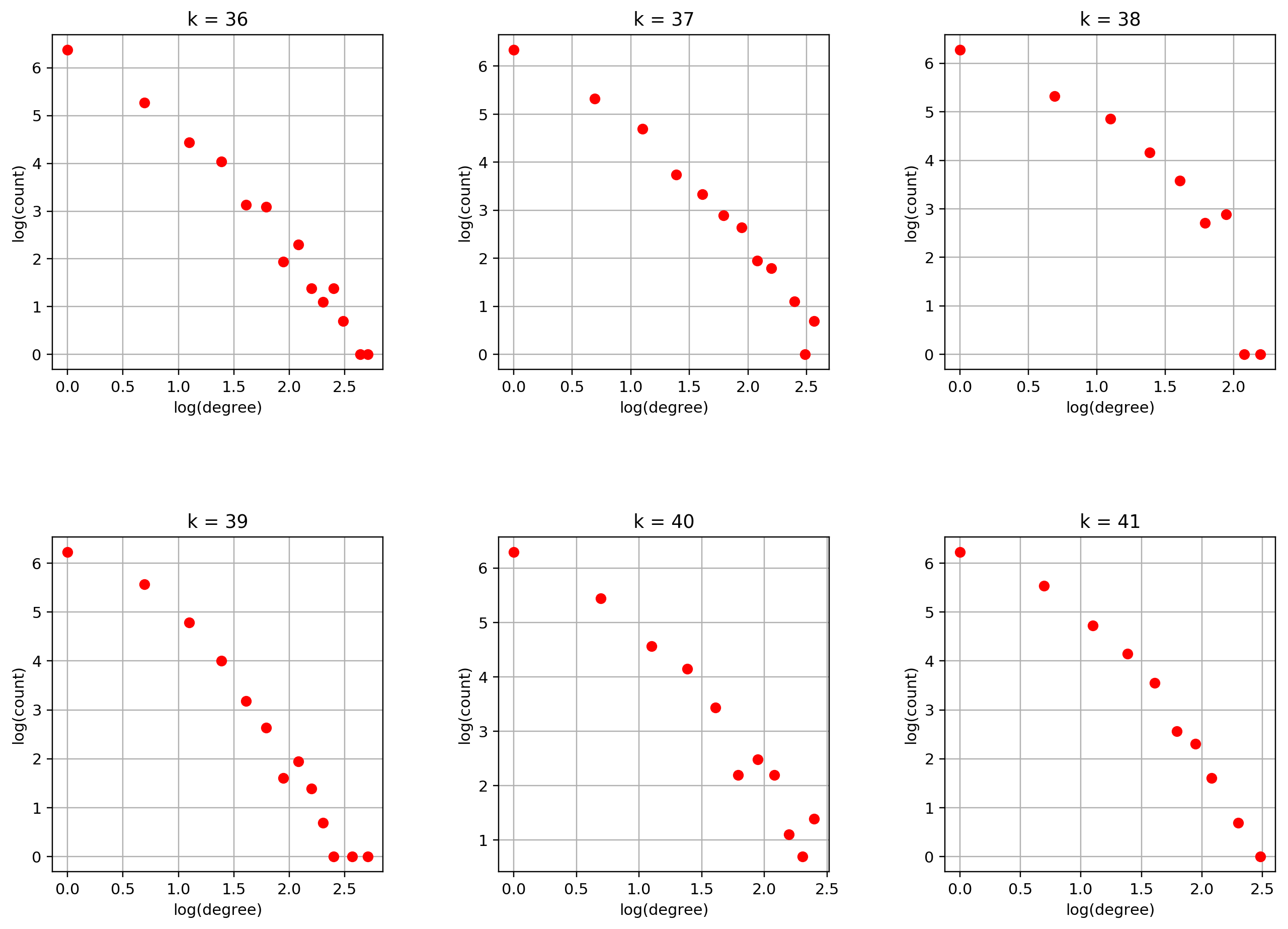}
\includegraphics[scale = 0.29]{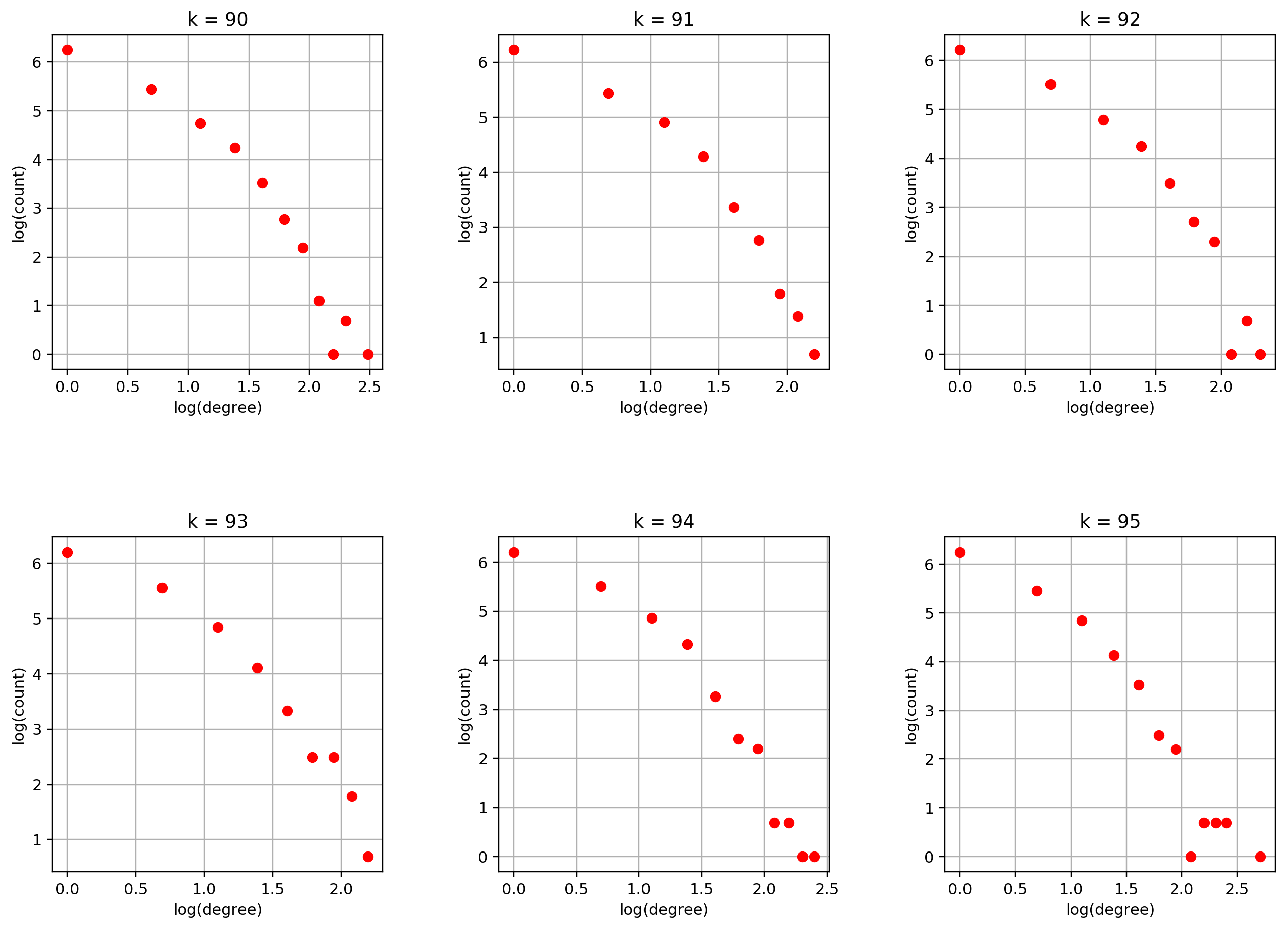}

According to the plots that are generated we can infer the 
following: 
\begin{itemize}
    \item For small values of $k$ this model gives rise to
    a scale free distribution when $k$ is even and for larger 
    values of $k$ the degree distribution follows the 
    \emph{stretched exponential distribution} in the log-scale.
    This in turn implies that in the normal scale the plot ensures
    that the number of nodes with high degree is fairly large as 
    compared to the \emph{scale free distribution} which implies that
    for large values of $k$ there are several nodes with high degree.
    
    \item The intuitive reasoning for the aforementioned result is
    that for small values of $k$ this model behaves basically similar 
    to the BA model. Whereas for larger values of $k$ we can observe 
    that once a new node gets attached to the already existing node the pulling power of the already existing node does not increase and the pulling power of the $k^{th}$-neighbor increases. This phenomenon creates some sort of an oscillation on the increase in the degree of the certain number of nodes which implies that this set of nodes experience enhancement of degree simultaneously. This explains the
    \emph{stretched exponential distribution} in the log-scale for larger values of $k$.

\end{itemize}

\subsection{Analysis of Model B}

The following are the plots that we 
obtain once the degree distributions of \emph{Model B} are generated
and based on these we infer the following 

\begin{itemize}
    \item As said earlier in this model the incoming node is attached to that node which is on the boundary of the $k$-neighborhood of the node that maximizes the objective function. 
    
    \item From the plots, it is clear that for both small and large values of $k$ we get a \emph{power law} distribution. This observation can be explained intuitively as follows: since the incoming node $x$ is getting attached to a node which is on the boundary of the $k$-neighborhood of a node $y$, the pulling power of node $y$ increases with time because it is defined as the number of edges on the boundary of the $k$-neighborhood of $y$ and hence in further iterations the node $y$ gets enhancement in its power.  
    
    \item This makes the model a generalization of the BA model which is quite similar to it.
    
\end{itemize}

\includegraphics[scale = 0.29]{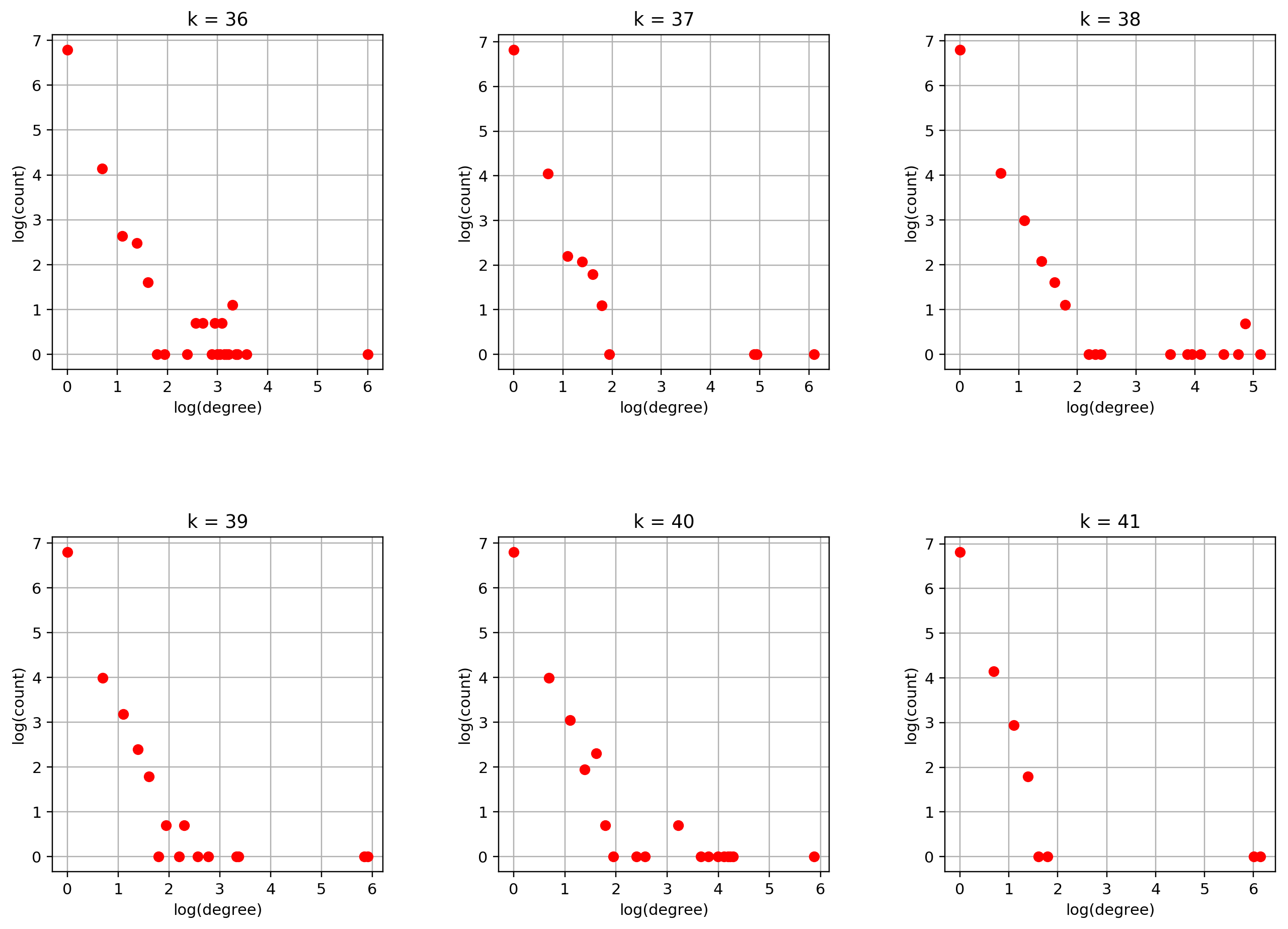}
\includegraphics[scale = 0.29]{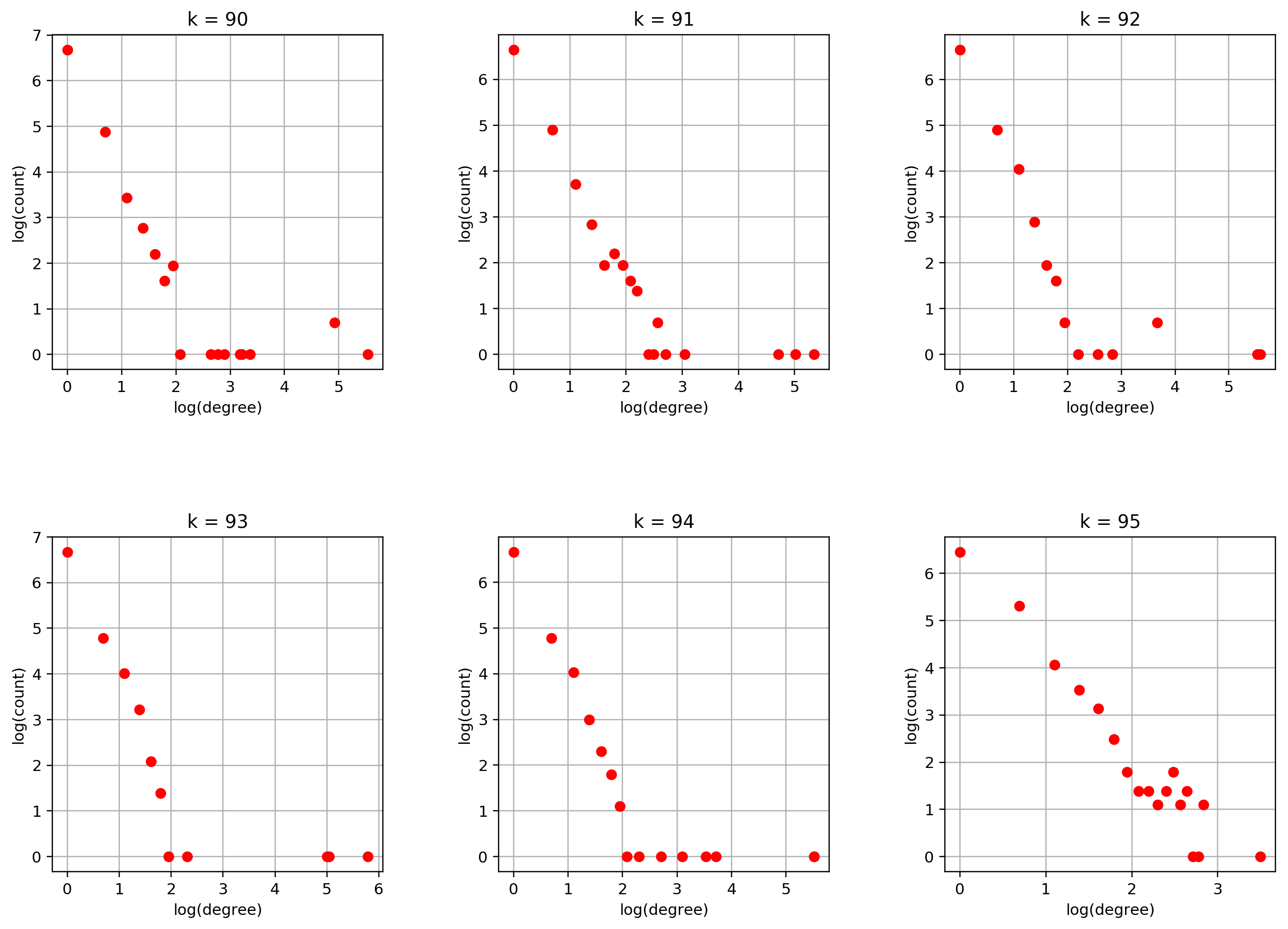}

\section{Comparison with Prior Models}
In this section we compare our models with the prior
generalizations of the BA model and also discuss how close
to reality these models can be considered.

\subsection{Comparison with DMS Model}
As mentioned before, the DMS model starts with an initial state
and the degree of the node is dependent on the initial state of the evolution process, a feature which
our models also possess. The DMS model achieves the power law
for large values of the degree whereas in our case the Model B
achieves power law even for small values of the degree. 

\subsection{Comparison with AKR Model}
In the AKR model the authors come up with a notion 
of a balance of the values present in a triad which is a collection of 
three nodes and the directed links among them. The dynamics of the model
tries to maintain the balance of the triads. This model, although a generalization of the BA model uses only \emph{local} modifications whereas our models are \emph{global} in the sense that each node looks around a $k$-neighborhood where $k$ can be fairly large. This allows our model to look at the global influence of the nodes which is not present in the AKR model.

\subsection{Comparison with SPSK Model}
In the SPSK Model, the evolution of the graph is based on some
operations done on the original graph called \emph{duplicate}, \emph{divergence}
and \emph{mutate}. This make the model very restrictive because a 
node can only get attached to the existing graph by performing a duplicate
operation which is basically the replication of the connection of an already 
existing node. Our models are not restrictive in that sense and a new node
is not forced to follow the topology of an already existing node and
is attached purely based on the pulling power of a particular node. Despite the
difference in nature of SPSK and our models we observe that the degree distribution
of this model and Model A proposed by us is quite similar and follows the stretched
exponential distribution.

\section{Real Life Implications of EvoCut}
One of the important questions regarding the models of evolution of
complex networks is how realistic are they. The BA model although
satisfactory is not considered to be very realistic. Thus there is 
definitely a need to come up with models which not only give rise to the power law distribution but are also realistic. In this section we justify how
our models can explain the realities of certain complex networks.
In the case of Model A, as we had observed that this model leads to
a degree distribution \cite{power-law networks} in which there are lots of nodes with large degrees. Although this doesn't follow the scale free property but explains the nature
of some real political networks. It is well known that \cite{Stretched exponential} 
the stretched exponential distribution describe very well the distributions of radio and light
emissions from galaxies, of country population sizes, of daily Forex US-Mark and
Franc-Mark price variations, of Vostok temperature variations and of citations of the most 
cited physicists in the world. This type of distribution may not be considered to be
explainable by the BA model and its generalizations which tend to result in a power law distribution,
but can be explained by Model A. \\
In the case of Model B we notice that since the model looks at a $k$-neighborhood
of a particular node to compute its pulling power this allows us to look
at a larger influence of a node as compared to the standard BA model. In fact in realistic
scenarios one can conceive of several situations in which when a new node
comes it doesn't get attached to the node with maximum pulling power but gets attached
to a node which is at a certain distance from the node. Consider the example 
of celebrities in real life networks, when a person considers himself as a follower
of a particular celebrity then he gets himself linked to another follower 
of the same celebrity rather than directly getting linked with the celebrity, and this
follower might as well not be one of the closest to the celebrity i.e. one gets attached to a \emph{fan club} rather than the celebrity himself. \\
We can also consider the example of any hierarchical ecosystem \cite{Hierarchical} present in
an organization like judiciary, police, banking system or a corporation in which
any customer is not allowed to connect with the highest authority in that system
but gets connected to a node that is lowest in that hierarchy. This notion
is being captured in the case of Model B, in the sense that the $k$-neighborhood of a node
basically creates a hierarchy  around that node considering that node as most powerful.
As the distance from the node increases we get down in the hierarchy, and hence 
when a new node comes under the influence of the $k$-neighborhood of a particular node at the $k^{th}$ level of the hierarchy. \\

\includegraphics[scale = 0.46]{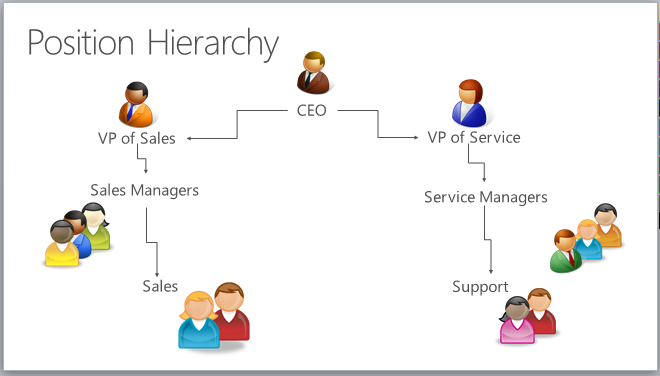}\\

Consider the aforementioned figure in which the CEO represents the node whose pulling power is being computed and the 1-neighborhood of it consists of VP of Sales and VP of Service, the 2-neighborhood consists of Sales Managers and Service Managers and the 3-neighborhood consists of Sales and Support. 
Thus when an new customer comes it is likely to get in touch with the 3-neighborhood rather than the CEO. Thus in several scenarios Model B gives a better picture of the evolution of the network and hence is much more realistic than the BA model.

\section{Conclusion and Future Work}
In this paper, we have presented a new generalization of the Barab\'asi-Albert (BA) model that is based on the neighborhood properties of nodes in the evolving graph. This generalization is substantially different from the previous generalizations of the BA model and one of its variant gives rise to the power law distribution and can model the growth of several real life networks which don't seem to be explainable by the BA model. The other variant of our Model, despite the fact doesn't give the scale free property models the growth of some other real life networks and gives rise to a stretched exponential distribution. We believe that our model is very powerful and further investigation into it can make it a stronger candidate in understanding the growth of several complex networks. \\
As part of future work it would interesting to get a mathematical proof of the  distribution generated by our models and also discover new interesting variants of our model.

\end{document}